\documentstyle[epsfig,amstex]{europhys}

\def\stars{\bigskip\centerline{***}\medskip}

\newif\ifboo \boofalse

\begin{document}
\euro{}{}{}{}
\Date{}
\shorttitle{R.~P.~SEAR, J.~A.~CUESTA: EMULSIFICATION FAILURE AND
BOSE-EINSTEIN CONDENSATION}
\title{What do emulsification failure and Bose-Einstein
condensation have in common?}
\author{Richard P.\ Sear\inst{1} and Jos\'e A.\ Cuesta\inst{2}}
\institute{
\inst{1}Department of Physics, University of Surrey,
	Guildford, Surrey GU2 7XH, United Kingdom,
	{\tt r.sear@@surrey.ac.uk} \\
\inst{2}Grupo Interdisciplinar de Sistemas Complicados (GISC), 
	Departamento de Matem\'a\-ti\-cas, Universidad Carlos III de
	Madrid, Avda.\ de la Universidad, 30, E-28911 -- Legan\'es, 
	Madrid, Spain, {\tt cuesta@@math.uc3m.es}}
\rec{}{}
\pacs{
\Pacs{03}{75Fi}{Phase coherent atomic ensembles; quantum condensation 
phenomena}
\Pacs{82}{70Kj}{Emulsions and suspensions}
\Pacs{05}{70Fh}{Phase transitions: general studies}
}

\maketitle

\begin{abstract}
Ideal bosons and classical ring polymers formed via self-assembly,
are known to
have the same partition function, and so analogous phase transitions.
In ring polymers, the analogue of Bose-Einstein
condensation occurs when a ring polymer
of macroscopic size appears.
We show that a transition of the same general form
occurs within a whole class of systems with self-assembly, and
illustrate it with the emulsification failure of a
microemulsion phase of water, oil and surfactant.
As with Bose-Einstein condensation,
the transition occurs even in the absence of interactions.
\end{abstract}

Bose-Einstein condensation (BEC) is a
textbook \cite{huang87}, but rather unusual, phase transition.
It occurs for {\em noninteracting} bosons, in contrast to more conventional
transitions such as that of the Ising model, which are driven by interactions.
Feynman \cite{feynman72} showed that the statistical mechanics of bosons
can be performed via what he called path integrals. These integrals
are, in turn, equivalent to integrals over the configurations of
ring polymers. Thus, the path integral formalism of Feynman implies
that noninteracting bosons and self-assembling, noninteracting ring
polymers have partition functions of exactly the same form.
Necessarily then, ring
polymers must undergo a phase transition precisely analogous to
BEC \cite{petschek86}. Here, we generalise this result to show
that there is a class of self-assembling systems which undergo
a phase transition analogous to BEC. This phase transition occurs
in the absence of interactions between the aggregates formed
by self-assembly. At BEC a condensate appears which is a macroscopic
number of bosons in a single state. In the analogous transition
in self-assembling systems, an aggregate of macroscopic size appears.
For the example considered here, a microemulsion, the macroscopic
aggregate is a bulk oil phase.

Self-assembling systems are systems in which the particles are not
immutable objects but are formed reversibly
\cite{israelachvili,gelbart94,schick94}.
Typically we have water, surfactant and sometimes oil.
The surfactant molecules then spontaneously assemble into micelles,
or coat and stabilise droplets of oil. Both the micelles and the
droplets are what we term aggregates.
For example, in a microemulsion,
equilibrium is obtained when the oil is dispersed in water as
oil droplets whose surfaces are coated with surfactant.
These droplets are more stable than just a single bulk oil phase
because with a single bulk phase there is no extensive oil-water interface,
and the amphiphilic surfactant molecules have the lowest energy
at this interface. The distribution of droplets changes with density and
temperature. A BEC-like transition occurs when, at equilibrium,
some of the oil exists as a macroscopic droplet, i.e., a bulk phase.
This transition is called {\em emulsification failure}
\cite{leaver94}.

Let us begin by outlining BEC in Feynman's language \cite{feynman72}.
The partition function of $N$ bosons
is given in terms of their density
matrix $\hat\rho$ by
\begin{equation}
Z_N(\beta,V)=\mbox{tr}_V\,\hat\rho, \quad \hat\rho\equiv e^{-\beta H_N},
\label{eq:partition}
\end{equation}
with $H_N$ the Hamiltonian operator, $\beta=1/kT$. The density matrix
of noninteracting identical bosons
can be written in terms of 1-particle density matrices,
$\hat\rho_1\big({\bf x}_i;{\bf x}'_i\big)$,
\begin{equation}
\hat\rho\big(\{{\bf x}_i\};\{{\bf x}'_i\}\big)=
\frac{1}{N!}\sum_{P\in\Pi_N}\prod_{i=1}^{N}
\hat\rho_1\big({\bf x}_i;P{\bf x}'_{P(i)}\big),
\label{eq:denssym}
\end{equation}
where
$\{{\bf x}_i\}\equiv\{{\bf x}_1,\dots,{\bf x}_N\}$ and
$\Pi_N$ is the set of permutations of $N$ elements
[$P:i\mapsto P(i)$].

Taking the trace over eq.~(\ref{eq:denssym}) and moving over
to the grand canonical ensemble, we obtain for
the grand partition function, $\Xi(\beta,z,V)$, and 
average density of bosons, $\phi$,
\begin{equation}
\ln\Xi=\sum_{n=1}^{\infty}\frac{h_nz^n}{n},\qquad
\phi=\frac{1}{V}\sum_{n=1}^{\infty}h_nz^n,
\label{eq:Q}
\end{equation}
with $z$ the activity of the bosons and
$h_n={\rm tr}_V\,\hat\rho_{_1}^n=Z_1(n\beta,V)$. It is 
straightforward to determine that
for a $d$-dimensional cubic box with periodic boundary condition
$Z_1(\beta,V)=[\vartheta(e^{-\pi/\ell^2})]^d$, where $\ell\equiv
V^{1/d}/\Lambda$ is the length of the box in units of the thermal
wavelength $\Lambda\equiv h\sqrt{\beta/2\pi m}$, 
and $\vartheta(q)\equiv 1+2\sum_{k=1}^{\infty}q^{k^2}$.
Thus, 
\begin{equation}
\phi=\frac{1}{V}\sum_{n=1}^{\infty}z^n\left[\vartheta(
e^{-\pi n/\ell^2})\right]^d.
\label{eq:npath}
\end{equation}

%
%

Remarkably,
eq.~(\ref{eq:Q}) also applies to self-assembling ring polymers,
provided $z$ and $\Lambda$ are properly redefined. To see this, note
that by definition \cite{feynman72}, the one-particle density matrix, 
$\hat\rho_1=\exp(-\beta H_1)$, is a solution of the equation
$\partial_{\beta}\hat\rho_1=-H_1\hat\rho_1=
(\hbar^2/2m)\nabla^2\hat\rho_1$.
This is just
a Fokker-Planck equation for a random walk in ``time'' $\beta$,
and with diffusion coefficient $\hbar^2/2m$.
Accordingly, $V^{-1}{\rm tr}_V\,\hat\rho_1^n(=h_n/nV)$ has an
alternative interpretation as the probability of a random
walk forming a cyclic path after $n$ ``time''-steps of length $\beta$.
The simplest model of a ring polymer of length $n$ is a 
cyclic random walk of $n$ steps; hence $h_n/nV$ can also
be regarded as the ratio of the partition function
of a ring of $n$ monomers, $Z_n^{\rm ring}$, to that of
a chain of $n+1$ monomers, $V\Lambda^{nd}$. With this
interpretation $\Lambda$ has the meaning 
of the average monomer length.
We conclude that $Z_n^{\rm ring}=\Lambda^{nd}h_n/n$.
For an ideal mixture of rings made up of monomers at an activity $z$,
the number density of rings of length $n$,
$\rho(n)=z^nZ_n^{\rm ring}/V$, and
the number density of monomers in rings of length $n$
is just $n$ times this, $(z\Lambda^d)^nh_n/V$.
If we absorb a factor of $\Lambda^d$ into $z$ then clearly the total
number density of monomers in ring polymers is given exactly by the
expression for the density of bosons in eq.~(\ref{eq:Q}).
This well-known mapping of ring polymers onto bosons
is called the ``classical isomorphism'' \cite{chandler81}. It
tells us that both systems have {\em completely equivalent}
equilibrium behaviour, and so
(i) the ring polymer system must undergo the analogue of
BEC, and (ii) ideal bosons are equivalent to a classical system with
self-assembling aggregates.

To locate the phase transition for bosons/ring polymers let
us consider an integer $n_1(V)$ such that $n_1\ll\ell^2$ 
but $n_1(V)\to\infty$ as $V\to\infty$. Then
$\vartheta(e^{-\pi n/\ell^2})=\ell n^{-1/2}+O(e^{-\pi\ell^2})$
for all $n\leq n_1(V)$, and so we can split the sum of 
eq.~(\ref{eq:npath}), yielding
\begin{equation}
\phi=\frac{1}{\Lambda^d}\sum_{n=1}^{n_1}\frac{z^n}{n^{d/2}}+
\frac{1}{V}\sum_{n=n_1}^{\infty}z^n\left[\vartheta(
e^{-\pi n/\ell^2})\right]^d+\epsilon(V),
\label{eq:nsplit}
\end{equation}
where $\epsilon(V)\to 0$ as $V\to\infty$.
If we fix $z<1$ and take the thermodynamic limit then the
first sum is less than its value at $z=1$,
$\phi_c=\Lambda^{-d}\sum_{n=1}^{\infty}n^{-d/2}$,
and the second sum is zero.
When $z>1$ the first sum is clearly divergent. 
This implies that all $\phi>\phi_c$
correspond to $z=1$. Careful analysis of eq.~(\ref{eq:nsplit})
shows \cite{sear00} that for $\phi>\phi_c$, the density in 
excess of $\phi_c$, i.e.,
$\phi-\phi_c$, comes from the second sum of eq.~(\ref{eq:nsplit}).
This second sum is the contribution of either a
macroscopic number of bosons in a single state (the ground state)
or a ring polymer of macroscopic size.

We now generalise the preceding theory to describe a whole
class of systems with self-assembly.
Let us consider monomers which can self-assemble into aggregates,
where these aggregates are noninteracting but are otherwise
arbitrary. The {\em internal} partition function
of the corresponding aggregates in the new model
will differ from that of ring polymers, but density of matter,
$\phi$, will again be given
in terms of the activity, $z$, through some function $\phi=G(z)$.
We will then have a BEC-like transition provided that
as $z$ runs over a range $0\leq z\leq z_c$, the new function 
takes only values in the limited range $0\leq G(z)\leq G(z_c)$, and
is divergent for $z>z_c$. Furthermore, this phase transition 
will again occur due to the appearance of macroscopic aggregates.

Let us denote by $s$ the (dimensionless) ``size'' (volume,
surface, length\dots) of aggregates, and assume
$s_0\leq s<\infty$. If $\rho(s)$ denotes the number density
of aggregates of size $s$, then the
free energy of an ideal mixture of these aggregates is
\begin{eqnarray}
\beta F/V&=&
\sum_{s\geq s_0}\rho(s)\left[\ln\rho(s)-1+f(s)+as\right]
\nonumber\\&&
+\rho_0[\ln\rho_0-1+f_0]
\label{eq:F}
\end{eqnarray}
where $f(s)$ is the internal free energy of aggregates of size $s$,
with the linear part, $as$, subtracted off. The second term must 
be included if aggregates for which $s=0$ are present
(for example, in microemulsions micelles are present which are pure
surfactant and do not contribute to $\phi$; see below). In this
term $\rho_0$ and $f_0$ are the number density and internal free
energy of these aggregates.
The sum in eq.~(\ref{eq:F}) may be replaced by an integral
with only very minor quantitative changes to the behaviour.
We will do so when we come to do explicit calculations for
microemulsions, for which $s$ may be treated as a continuous
variable. Note that
ring polymers conform to eq.~(\ref{eq:F}) if $s$ denotes the
number of monomers (i.e.\ the dimensionless length), $s_0=1$
and $f(s)+as=-\ln[Z^{\rm ring}(s)/V]$ (the $s=0$ term is
obviously absent).

A microemulsion is an isotropic equilibrium phase
composed of water, oil and a surfactant \cite{strey94,gelbart94}.
If there is, say,
much less oil than water then the microemulsion
consists of droplets of oil, nanometers across, coated with the
surfactant and dispersed in the water. Without the
surfactant, at equilibrium we would have bulk oil 
and water phases, but as surfactant prefers to lie in the 
water-oil interface, it stabilises
the oil droplets dispersed in the water.
For microemulsions the sum of eq.~(\ref{eq:F}) is over droplets
of different sizes; here
$s$ is the reduced volume of a droplet and $s_0$ is the size of the
smallest possible droplet. Micelles, aggregates of
pure surfactant, are also present in
microemulsions, and as they are pure surfactant they do not contribute
to the oil density $\phi$. In eq.~(\ref{eq:F}) applied to microemulsions,
the second term on the right hand side accounts for micelles;
$\rho_0$ is the density of micelles.
The surfactant is assumed
to be restricted to the surface of the droplets and to micelles.

At equilibrium the droplet distribution $\rho(s)$ is given by the
minimum of the free energy, subject to the constraints
that the total fraction of oil,
\begin{equation}
\phi=\sum_{s\geq s_0}s\rho(s),
\label{eq:matter}
\end{equation}
is fixed, and also
that the total amount of surfactant is fixed.
Fixing the (reduced) surfactant density $\xi$ requires constraining the
sum of the micelle density, $\rho_0$, and the $s^{2/3}$ moment of $\rho(s)$.
The latter is proportional to the surface area of the droplets per unit
volume. This constraint is a single specific example of constrained densities 
$\xi_i$ of the form
\begin{equation}
\xi_i=c_i\rho_0+\sum_{s\geq s_0}w_i(s)\rho(s),\quad i=1,\dots,\nu,
\label{eq:constraints}
\end{equation}
where $c_i$ are constants,
$\rho_0$ is a density of a species which contributes to $\xi_i$
but not to $\phi$, and $w_i(s)$ are weight functions.
Equations (\ref{eq:F}), (\ref{eq:matter}) and (\ref{eq:constraints})
define our model of a system with self-assembly.


Minimising the general free energy, eq.~(\ref{eq:F}),
at constant $\phi$, eq.~(\ref{eq:matter}),
under the constraints (\ref{eq:constraints}), we obtain the
equilibrium densities
\begin{equation}
\rho(s)=e^{-f(s)-\sum_i\lambda_iw_i(s)-
\lambda s}, \ \
\rho_0=e^{-f_0-\sum_i\lambda_ic_i},
\label{eq:distribution}
\end{equation}
where $\lambda=a-\beta\mu$, $\mu$ being the chemical potential 
corresponding to the reduced 
density $\phi$, and $\lambda_i$ are the Lagrange multipliers
corresponding to the constraints (\ref{eq:constraints}) (and
which are analogous to chemical potentials associated with the
fixed densities $\xi_i$). 

Now, if either $f(s)$ or any of the $w_i(s)$
grows faster than $s$ as $s\to\infty$, then the sum in
eq.~(\ref{eq:matter}) will converge for {\em any} value of
$\lambda$. Furthermore, for {\em any} set of densities $\{\phi,
\xi_i\}$ there will always be a set $\{\lambda,\lambda_i \}$,
which will solve eqs.~(\ref{eq:constraints}) and (\ref{eq:matter}).
This means that the free energy will be a smooth function
of the densities and no phase transition will occur. Therefore,
for a phase transition to occur, it is necessary that (a)
$f(s),w_i(s)=o(s)$ as $s\to\infty$.
If this condition holds then the sum
in eq.~(\ref{eq:matter}) is convergent for $\lambda>0$ and
divergent for $\lambda<0$. The second condition concerns
the limit $\lambda\to 0^+$. If the sum in eq.~(\ref{eq:matter}) 
diverges in this limit, then again any set of densities will be
reachable and no phase transition will occur. Thus, for the 
transition to appear it is necessary that (b) the sum in
eq.~(\ref{eq:matter}) is convergent for $\lambda=0$.

If our model fulfills conditions (a) and (b) we have that
the density $\phi$ increases to a finite value and then suddenly
diverges, just as for bosons. Just as with bosons, to study the
transition we need to consider explicitly the volume dependence.
For this we need to consider a free energy of aggregates
which depends on $V$, i.e., $\ln\rho(s)-1+f(s;V)+as$. Now, $f(s;V)$
has to be such that $f(s;V)\to f(s)$ as $V\to\infty$,
but the convergence cannot be uniform: intuitively, if $s$ is
``small'' compared to $V$, then $f(s;V)\approx f(s)$, but
if $s$ is ``large'', then $f(s;V)$ will differ markedly from $f(s)$.
Thus we can always choose a size $s_1(V)$ such that
$f(s<s_1;V)\approx f(s)$ and that 
$s_1(V)\to\infty$ when $V\to\infty$.

There are two possible scenarios
depending on whether $f(s;V)$ violates either condition 
(a) or condition (b). For example, condition (a) is violated
when compact aggregates form and the box limits
the maximum size of an aggregate.
This translates into a sharp
increase of $f(s;V)$ at a certain macroscopic size. Case
(b) is what we have met for ring polymers, for which 
$f(s;V)\sim\ln(Vs)$ as $s\to\infty$. Although there are small
differences between the two cases \cite{sear00}, they do not
affect the dominant terms in the thermodynamic limit.
We report the details of the analysis elsewhere \cite{sear00}, and
limit ourselves to describing the resulting phase behaviour.
If $\phi<\phi_c$, eqs.~(\ref{eq:matter})
and (\ref{eq:constraints}) can be solved for 
$\lambda$ and $\lambda_i$. This yields the densities,
eq.~(\ref{eq:distribution}),
as well as the free energy
\begin{equation}
\beta F/V=-\rho-\sum_{i=1}^{\nu}\lambda_{i}\xi_i+a\phi,
\label{eq:Fnotcondens}
\end{equation}
with $\rho=\rho_0+\sum_s\rho(s)$.
The phase transition occurs at a density $\phi_c$, which
is given by setting $\lambda=0$ in eq.~(\ref{eq:distribution})
for $\rho$ and inserting the result in eq.~(\ref{eq:matter}).
When $\phi\geq\phi_c$, the matter in excess of $\phi_c$, $\phi-\phi_c$
forms macroscopically big aggregates (the {\em condensate}),
and $\lambda=0$. The density of non-macroscopic aggregates is then
given by eq.~(\ref{eq:distribution})
with $\lambda=0$ and $\lambda_i=
\lambda_{i,c}$, where $\lambda_{i,c}$ are the solution to
eqs.~(\ref{eq:constraints})
for $\lambda=0$. In particular, as $f(s)$ increases sublinearly
then whenever macroscopic aggregates are present, $\lambda=0$ and
the size distribution of aggregates always decays {\em slower} than
exponentially.
Finally, in this regime the free energy has the same form as in
eq.~(\ref{eq:Fnotcondens}), but with $\rho$ and $\lambda_i$
replaced by $\rho_c$ and $\lambda_{i,c}$, the corresponding
values for $\lambda=0$.

Returning to our example of microemulsions, we will
explicitly calculate a phase diagram.
As mentioned before, $s$ is proportional to the volume.
The internal free energy of a droplet has a linear
term, $as$, plus $f(s)$, which comes
from the droplet's surface. The surface of the drop can be modeled
using a surface tension term plus an elastic free energy of the form
introduced by Helfrich \cite{helfrich73,strey94}, which
accounts for a preferred curvature of the surfactant layer at the
surface of the droplet. The details are given in Ref.~\cite{sear00},
but as we might expect there are
three contributions arising from the surface tension (proportional
to $s^{2/3}$), the mean curvature (proportional to $s^{1/3}$)
and the Gaussian curvature (a constant)\footnote{\label{fnote} 
Reiss and coworkers \cite{reiss96} have looked
carefully at the function $f(s)$ for microemulsions and found in
addition to the terms in eq.\ (\ref{eq:femul}) a term logarithmic in 
$s$. This term essentially accounts for fluctuations of the center of mass of
the droplet. Adding such a term to eq.\ (\ref{eq:femul}) results in a $f(s)$ 
which still satisfies condition (a) for a transition, and thus does 
not alter the nature of the phase behaviour. It will however shift 
the phase boundary.}. Thus we take for $f(s)$
\begin{equation}
f(s)=a_0-a_1s^{1/3}+a_2s^{2/3},
\label{eq:femul}
\end{equation}
consistent with requirement (a) above; $a_1>0$, the signs of
$a_0$ and $a_2$ are not fixed. This expression for $f(s)$
may be inserted in the expression for the free energy, eq.~(\ref{eq:F}),
and the sum replaced by an integral;
$s$ may be any volume in the range $s_0\leq s<\infty$. There is
a single constraint, that on the surfactant density
\begin{equation}
\xi=c_0\rho_0+c_1\int_{s_0}^{\infty}ds\,s^{2/3}\rho(s)
\label{eq:xiemu}
\end{equation}
($c_0$ and $c_1$ are simply geometric factors).
Minimising the free energy subject to fixed $\phi$ and $\xi$
yields the droplet size distribution $\rho(s)=\exp\{-a_0+
a_1s^{1/3}-(a_2+\lambda_2c_1)s^{2/3}-\lambda s\}$, and
density of micelles $\rho_0=e^{-f_0-c_0\lambda_2}$.
If $a_2+\lambda_2c_1>0$ the requirement (b) on $f(s)$
also holds and we will have a finite $\phi_c$ at which a 
transition occurs. At $\phi_c$ a macroscopic
droplet forms, which is nothing other than
a bulk oil phase. So, a bulk phase of excess oil
coexisting with the microemulsion phase has formed. This is 
the so-called {\em emulsification failure} \cite{leaver94}.

\begin{figure}
\begin{center}
\epsfig{file=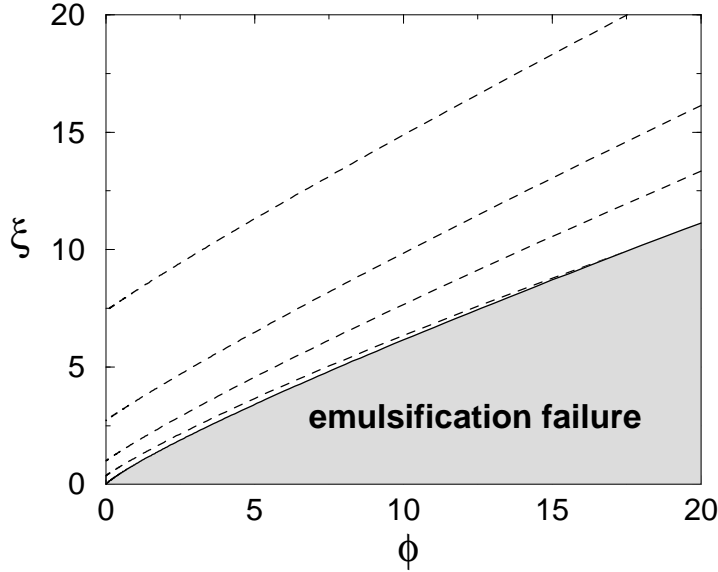,width=3.7in}
\end{center}
\caption{
Amount of oil in droplets, $\phi$, vs.\ amount of surfactant, $\xi$.
In the shaded region a bulk oil phase coexists with the
microemulsion (emulsification failure).
The solid curve ($\lambda=0$) signals the phase transition.
The dashed curves are at constant micelle density $\rho_0=e^{-\lambda_2}$.
{}From bottom to top $\lambda_2=-1,0,1,2$.
}
\label{figphixi}
\end{figure}

To illustrate emulsification failure we take some simple
and rather arbitrary values for the parameters of the free energy,
namely $a=a_0=a_2=f_0=0$, $a_1=c_0=c_1=s_0=1$, and calculate the
phase diagram, fig.~\ref{figphixi}. At fixed surfactant density $\xi$
there is a maximum oil density $\phi_c$ beyond which the microemulsion
phase coexists with a bulk oil phase. This maximum density increases
as the surfactant density increases.

Finally we briefly comment on interaggregate interactions.
Their presence will of course affect any transition which
is driven by the mechanism outlined in this Letter. But in
addition, Zhang {\it et al.} \cite{zhang99} report
evidence of a transition like that found here for a system
for which $f(s)=$const.. There, the {\em interactions} between the aggregates
lead to a distribution of aggregates of the form of
eq.~(\ref{eq:distribution}) but with $f(s)$ depending on the density
as a result of the interactions, not just arising from an internal
free energy \cite{zhang99,cuesta99,blaak00}. If this $f(s)$
satisfies conditions (a) and (b) there will be a transition
induced by the interactions.

There is a class of systems in which aggregates of all sizes
can reversibly self-assemble,
and in which, without interaggregate interactions, there is a phase
transition. This phase transition occurs when an infinite aggregate or
aggregates form. It is analogous to the Bose-Einstein
condensation of ideal bosons: in neither case are there interactions,
and the macroscopic aggregates we find play the same role as the
condensate of a macroscopic number of bosons.
This analogy for the specific case when the aggregates
are ring polymers is implied by Feynman's path integral
approach for bosons \cite{feynman72,chandler81,petschek86}.
We have shown that the analogy applies to a whole
class of self-assembling systems with partition functions
which are qualitatively similar, although of a different functional 
form, to that of ideal bosons/ring polymers.
This class is defined by the requirement that
the internal free energy of an aggregate contains a linear term, $as$,
plus others which are sublinear and ensure that the density
is finite when the chemical potential $\beta\mu\to a^{-}$.
We outlined the behaviour of one example here, that of microemulsions.
Another example is that of the formation of the lamellar phase,
a stack of effectively infinite bilayers of surfactant, which
can coexist with small discs of bilayer \cite{boden87}.
This transition, as well as emulsification failure, will be
affected by the presence of interactions between the aggregates.
We suggest that this effect will be analogous to that of interactions on
the $\lambda$ transition of liquid $^4$He \cite{huang87},
thought to be
Bose-Einstein condensation modified by the presence of interactions
between the He atoms.

\stars

This work is funded by projects HB1998-0008 (Ministerio de
Educaci\'on y Cultura and The British Council) and BFM2000-0004
(Direcci\'on General de Investigaci\'on, Ministerio de Ciencia
y Tecnolog\'{\i}a, Spain).


\begin{thebibliography}{99}

\bibitem{huang87} HUANG K., {\em Statistical Mechanics}, 2nd
        ed.\ (Wiley, New York, 1987).
\bibitem{feynman72} FEYNMAN R.\ P., {\em Statistical Mechanics:
        A Set of Lectures} (Addison Wesley, Redwood City, 1972).
\bibitem{petschek86} PETSCHEK R.\ G., PFEUTY P.\ and WHEELER J.\ C.,
        {\em Phys.\ Rev.\ A} {\bf 34}, 2391 (1986).
\bibitem{israelachvili} ISRAELACHVILI J.\ N.,
        {\it Intermolecular and Surface Forces}
        (Academic Press, London, 1992).
\bibitem{gelbart94} 
	GELBART W.\ M., BEN-SHAUL A.\ and ROUX D., eds.,
	{\em Micelles, Membranes, Microemulsions and Monolayers}
        (Springer-Verlag, New York, 1994).
\bibitem{schick94} SCHICK M.\ and GOMPPER G., in
	{\it Phase Transitions and Critical Phenomena Vol 16},
	edited by C.\ DOMB and J.\ L.\ LEBOWITZ
	(Academic Press, London, 1994).
\bibitem{leaver94} LEAVER M.\ S., OLSSON U., WENNERSTROM H.\ and 
	STREY R., {\em J.\ Phys.\ II} {\bf 4}, 515 (1994); VOLLMER D., 
	STREY R.\ and VOLLMER J., {\em J.\ Chem.\ Phys.}\ {\bf 107}, 
	3619 (1997).
\bibitem{chandler81} CHANDLER D.\ and WOLYNES P.\ G., {\em J.\ Chem.\
        Phys.}\ {\bf 74}, 4078 (1981); CEPERLEY D.\ M., {\em Rev.\ Mod.\ 
	Phys.}\ {\bf 67}, 279 (1995).
\bibitem{sear00} CUESTA J.\ A.\ and SEAR R.\ P., to be published.
\bibitem{strey94} STREY R., {\em Colloid Polymer Sci.}\ {\bf 272}, 1005
        (1994).
\bibitem{helfrich73} HELFRICH W., {\em Z.\ Naturforsch.}\ {\bf 28a}, 693
        (1973).
\bibitem{reiss96} REISS H., KEGEL W.\ K., GROENEWOLD J.,
	{\em Ber.\ Bunsen Phys.\ Chem.}\ {\bf 100}, 279 (1996); 
	KEGEL W.\ K., REISS H., {\em ibid.}\ {\bf 100}, 300 (1996).
\bibitem{zhang99} ZHANG J., BLAAK R., TRIZAC E., CUESTA J.\ A.\
        and FRENKEL D., {\em J.\ Chem.\ Phys.}\ {\bf 110}, 5318 (1999).
\bibitem{cuesta99} CUESTA J.\ A.\ and SEAR R.\ P.,
        {\em Eur.\ Phys.\ J.\ B} {\bf 8}, 233 (1999).
\bibitem{blaak00} BLAAK R., {\em J.\ Chem.\ Phys.}\ {\bf 112}, 9041 (2000).
\bibitem{boden87} BODEN N., CORNE S.\ A.\ and JOLLEY K.\ W.,
        {\em J.\ Phys.\ Chem.}\ {\bf 91} 4092 (1987); ZEMB TH., DUBOIS M.,
	DEM\'{e} B.\ and GULIK-KRZYWICKI TH., {\em Science} {\bf 283}, 816 
	(1999).

\end{thebibliography}
\end{document}